\documentclass[journal,twoside,web]{ieeecolor}
\usepackage{generic}
\usepackage{cite}
\usepackage{amsmath,amssymb,amsfonts}
\usepackage{graphicx}
\usepackage{textcomp}
\usepackage{multirow}
\usepackage{amsmath}
\usepackage{stfloats}
\usepackage{color}
\usepackage{booktabs}

\usepackage{mathrsfs}

\usepackage{easyReview}
\usepackage[switch]{lineno}    
\usepackage[ruled]{algorithm2e}

\makeatletter
\renewcommand{\maketag@@@}[1]{\hbox{\m@th\normalsize\normalfont#1}}%
\makeatother

\def\BibTeX{{\rm B\kern-.05em{\sc i\kern-.025em b}\kern-.08em
    T\kern-.1667em\lower.7ex\hbox{E}\kern-.125emX}}
\begin{document}
\title{Eye tracking guided deep multiple instance learning with dual cross-attention for fundus disease detection}

\author{Hongyang Jiang, Jingqi Huang, Chen Tang, Xiaoqing Zhang, Mengdi Gao, Jiang Liu  
\thanks{This work was supported in part by General Program of National Natural Science Foundation of China under Grant 82272086, in part by Shenzhen Science and Technology Program under Grant KQTD20180412181221912 and Grant JCYJ20200109140603831. \textit{(Corresponding author:  Mengdi Gao; Jiang Liu.)}}
\thanks{Hongyang Jiang, Jingqi Huang, Xiaoqing Zhang and Jiang Liu are with the Department of Computer Science and Engineering, Southern University of Science and Technology, Shenzhen 518055, China (e-mail: jianghy3@sustech.edu.cn; huangjq@mail.sustech.edu.cn; 11930927@mail.sustech.edu.cn; liuj@sustech.edu.cn).}
\thanks{Chen Tang is with the School of Ophthalmology and Optometry and Eye Hospital, Wenzhou Medical University, Wenzhou 325027, China (e-mail: tangchen419@163.com)}
\thanks{Mengdi Gao are with the Department of Biomedical Engineering, College of Future Technology, Peking University, Beijing 100871, China (e-mail: gaomengdi@pku.edu.cn).}
}
\maketitle

\begin{abstract}
Deep neural networks (DNNs) have promoted the development of computer aided diagnosis (CAD) systems for fundus diseases, helping ophthalmologists reduce missed diagnosis and misdiagnosis rate. 
However, the majority of CAD systems are data-driven but lack of medical prior knowledge which can be performance-friendly. 
In this regard, we innovatively proposed a human-in-the-loop (HITL) CAD system by leveraging ophthalmologists' eye-tracking information, which is more efficient and accurate. 
Concretely, the HITL CAD system was implemented on the multiple instance learning (MIL), where eye-tracking gaze maps were beneficial to cherry-pick diagnosis-related instances.
Furthermore, the dual-cross-attention MIL (DCAMIL) network was utilized to curb the adverse effects of noisy instances.
Meanwhile, both sequence augmentation module and domain adversarial module were introduced to enrich and standardize instances in the training bag, respectively, thereby enhancing the robustness of our method.
We conduct comparative experiments on our newly-constructed datasets (namely, AMD-Gaze and DR-Gaze), respectively for the AMD and early DR detection.
Rigorous experiments demonstrate the feasibility of our HITL CAD system and the superiority of the proposed DCAMIL, fully exploring the ophthalmologists’ eye-tracking information. These investigations indicate that physicians’ gaze maps, as medical prior knowledge, is potential to contribute to the CAD systems of clinical diseases.
\end{abstract}


\section{Introduction}
\label{sec:introduction}

\IEEEPARstart{E}{ye} 
care has become an ongoing concern for worldwide researchers. Since 2020, there has been more than one billion people suffering vision impairment or blindness from potentially preventable or correctable causes \cite{assi2021global}. Some computer aided diagnosis (CAD) screening systems based on DNNs have been developed for some vision-threatening fundus diseases, such as diabetic retinopathy (DR) and aged-related macular degeneration (AMD). As recent studies report, the prevalence of diabetes around the world will reach 592 million people by 2035 \cite{guariguata2014global}, with one-third affected by the complicating disease DR \cite{lee2015epidemiology}. Besides, the number of AMD patients worldwide is estimated to reach around 200 million in 2040 \cite{wong2014global}. These serious situations impose a heavy burden on the early detection of fundus diseases.

Color fundus photographs has become the most commonly-used and significant diagnostic resource of fundus diseases. With the booming of the well-labeled fundus images, DNNs-based screening systems have emerged, gradually achieving comparable performance with ophthalmologists. Generally, data annotation are optional for deep learning (DL) models, i.e., image-level, box-level, and pixel-level. Therefore, researchers devise models from two aspects accordingly. 
First, some high-quality network structures are proposed via utilizing qualitative image-level labels. However, good performance and reasonable visualization results, such as class activate map (CAM) \cite{zhou2016learning}, are highly-dependent on massive-scale annotations. 
Second, researchers attempt to design models based on refined box-level or pixel-level labels, to accurately locate or segment lesions, such as microaneurysm (MA), haemorrhage (HM), soft or hard exudate (EX), and drusen. Although they can provide rational inference basis, the labeling process, which is time-consuming and effortful, especially for the pixel-level labeling, restricts the sustainable research.

The critical issue is that we should rethink how to boost the performance of CAD systems without additional labeling burden. This paper attempts to explore more medical prior knowledge during the diagnosis. Instead of heavy manual data annotation, eye movement information of an ophthalmologist during reading fundus images can be collected conveniently. Meanwhile, evidence-based medicine demonstrates that the image content where ophthalmologists pay more attention can support the diagnostic conclusion. Therefore, the eye-tracker which is utilized to capture eye movement information becomes an optimal auxiliary equipment during the diagnosis.

Eye-tracker has been utilized in many human health related researches, such as psychology \cite{mele2012gaze}, neuroscience \cite{popa2015reading}, and other diseases assisted diagnostic systems \cite{jiang2023eye,khosravan2019collaborative}. In particular, the eye-tracker has been employed in different phases of CAD development, including image annotation \cite{stember2019eye} and model design \cite{wang2022follow}. As proven in \cite{karargyris2021creation}, compared with manual pixel-level annotation, eye-tracking is a low-cost rough annotation method for medical image. Moreover, eye-tracking is a kind of imperceptive data acquisition behavior during the diagnosis of ophthalmologists, without spending non-diagnostic time additionally, which relieves the burden of data annotation. 

In our study, the eye-tracker is considered as a necessary device in the decision-making process of CAD. Using the generated ophthalmologists’ gaze maps from the eye-tracker, we innovatively propose a human-computer cooperation aided diagnosis framework. The main contributions of this paper are summarized as follows:
\begin{itemize}
	\item We propose a human-in-the-loop (HITL) CAD system for fundus disease detection with eye movement of ophthalmologists as medical prior knowledge. 
	\item HITL CAD system is implemented through the multiple instances learning and ophthalmologists’ gaze maps are initially utilized to screen instances which facilitates the correct diagnosis. Additionally, we propose a novel dual-cross-attention MIL (DCAMIL) model with the contrast learning regularization, suppressing the noisy instances. 
	\item Furthermore, we introduce the sequence augmentation module and domain adversarial network to enrich and standardize instances in the training bag, thereby enhancing the robustness of our HITL CAD system.
	\item We independently construct DR-Gaze and AMD-Gaze datasets, including fundus photographs and corresponding gaze maps. Both datasets validate the effectiveness and superiority to the state-of-the-art of the proposed DCAMIL in the fundus disease diagnostic scenario.
\end{itemize}

\section{Background}
To improve the detection accuracy of vision-threatening fundus diseases, our study takes ophthalmologists’ gaze maps recorded from the eye-tracker as requisite medical prior knowledge to make a HITL CAD system for DR and AMD detection. In this section, several related works are introduced. First, we review recent fundus disease detection methods, especially for DR and AMD. Then, we introduce some relative eye-tracking researches in CAD. Finally, we present the MIL technology applied in medical image analysis.

\subsection{Fundus disease detection based on DL}
The development of deep convolutional neural network (CNN) in computer vision has facilitated the fundus disease detection based on color fundus image. A variety of advanced DL network structures have been employed for the research of major fundus disease screening system, among which DR and AMD are of great concern. At present, there are many publicly available datasets with image-level or pixel-level annotation used for DR and AMD detection research \cite{khan2021global}. According to DR grading criteria \cite{wilkinson2003proposed}, researchers proposed several DR detection approaches, such as binary classification \cite{jiang2019interpretable,quellec2017deep}, multi-classification \cite{li2019diagnostic} and lesion-based classification \cite{huang2021lesion}. For instance, Liu et al. \cite{liu2019referable} proposed a multiple weighted paths CNN (WP-CNN) to distinguish referable DR and non-referable DR, acquiring better performance than pretrained ResNet, SeNet and DenseNet architectures. Li et al. \cite{li2019canet} presented a cross-disease attention network (CANet) to jointly grade DR and DME, and explored the individual diseases and also the internal relationship between two diseases. Guo et al. \cite{guo2019seg} proposed a multi-lesion segmentation model (L-Seg) that fusing multi-scale feature, to detect DR related lesions, i.e., MAs, HMs, soft and hard EXs. 

As for AMD detection, researches have conducted DL-based algorithm researches on AMD type classification (i.e., wet and dry AMD) and severity grading \cite{sengupta2020ophthalmic}. Tan et al. \cite{tan2018age} proposed a 14-layer CNN to detect AMD abnormal, including early, intermediate AMD, geographic atrophy and wet AMD. Chen et al. \cite{chen2019multi} developed a multi-task DL model based on InceptionV3 block for AMD classification of the AREDS severity scale \cite{davis2005age}. Liu et al. \cite{liu2019deepamd} applied several deep learning models into a MIL framework with two-stage to detect small lesions in the early stage of AMD.

\subsection{Eye-tracking research in CAD}
In recent years, some researchers have taken the eye-tracking as auxiliary information during the CAD design, which is reflected in two aspects. First, the eye-tracking is applied in automatic image annotation. Stember et al. \cite{stember2019eye} proposed a eye-tracking-based dynamic annotation method for lesions and organs on multimodal medical image, such as CT and MRI. Besides, Stember et al. \cite{stember2020integrating} also used eye-tracking and speech information together to label abnormal regions in medical images. Second, the clinical expert’s gaze map generated by the eye-tracker can serve as an alternative weakly supervised label to guide the training of DL models. Based on radiologists’ gaze data, Karargyris et al. \cite{karargyris2021creation} proposed a U-Net based multi-head model for synchronously predicting experts' attention and classification results about congestive heart failure (CHF), pneumonia and normal. Wang S et al. \cite{wang2022follow} proposed a gaze-guided attention model to grade the osteoarthritis into four categories based on X-ray images, which obtained better interpretability and classification performance.

\subsection{Multiple instance learning (MIL) in medical image analysis}
MIL is a weak supervised method, that has been widely used in medical image analysis with coarse annotation. In most circumstances, the annotation work of medical images is both professional and cumbersome, so some locally-annotated images have become the second choice for researchers. Presently, many MIL-based DL models have been proposed in medical image analysis scenarios, including cancer classification of different tissues in histopathology images \cite{yao2020whole}, lesions detection and cancer analysis in magnetic resonance and imaging (MRI) and computed tomography (CT) images \cite{qiu2021predicting,chen2022lung}, retinal disease recognition in optical coherence tomography (OCT) and color fundus images \cite{mueller2022multiple,wang2023deep} and so on. For instance, Sudharshan et. al. \cite{sudharshan2019multiple} compared several pioneering MIL methods on the public BreakHis dataset for breast tumor classification, and experiments demonstrated that the non-parametric MIL approach acquire the best results. Han et. al. \cite{han2020accurate} proposed an attention-based deep 3D MIL method based on chest CT images for achieving accurate and interpretable screening of COVID-9. Zhu et. al. \cite{zhu2021dual} presented a dual attention deep MIL network using structural MRI for the early diagnosis of Alzheimer's disease (AD) and its prodromal stage mild cognitive impairment (MCI). Moreover, Yu et. al. \cite{yu2021mil} proposed a MIL based head, which can be attached to the Vision Transformer, to enhance the model performance for fundus disease classification. 
\begin{figure*}[!t]
	\centering
	\includegraphics[width=2.0\columnwidth]{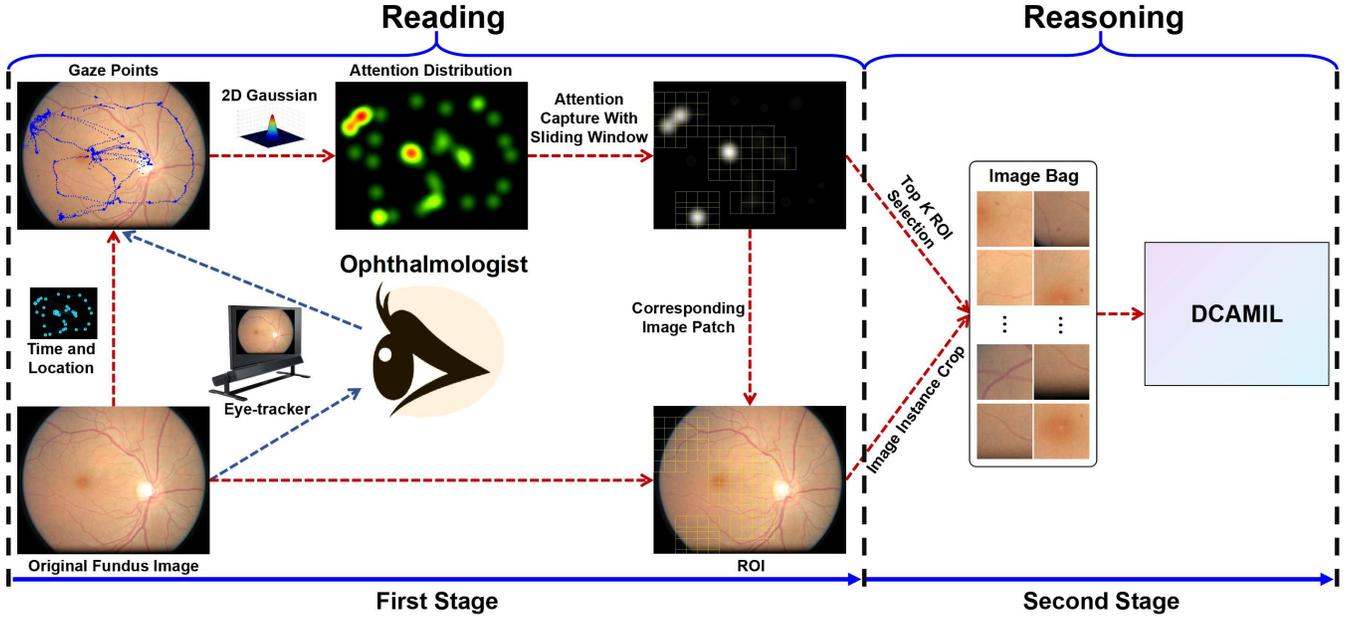}
	\vspace{-0.5em}
	\caption{HITL CAD system for fundus diseases.}
	\label{fig:HITL}
	\vspace{0.0em}
\end{figure*}

\section{Methodology}
In this section, we first introduce the preliminaries for the diagnosis of fundus diseases based on the eye-tracker. Then, we present the problem formulation and provide an overall solution framework. Finally, we analyze and discuss the details of each technical component in the proposed method.

\subsection{Diagnosis scenario description}
To fully excavate clinicians’ prior knowledge, we construct a HITL fundus disease diagnosis scenario through the eye-tracker. Specifically, we innovatively introduce an eye-tracker to record ophthalmologists’ eye-tracking information during their diagnosis for fundus diseases. According to the principles of evidence-based medicine \cite{sackett1996evidence}, the diagnostic process for ophthalmologists is a process of finding evidence to support their decision-making. Therefore, we devise an eye-tracking guided HITL CAD system for fundus diseases, which is illustrated in \textcolor{blue}{Fig \ref{fig:HITL}}. The HITL CAD system consists of two stages, i.e., reading and reasoning. In the reading stage, ophthalmologists need to sit in front of a 24-inch screen conducting routine fundus image reading, and the image size is standardized to $800\times800$. More specifically, ophthalmologists’ dynamic fixation points \( \left\{\left.(x, y)\right| 0 \leq x, y \leqslant 799; x, y \in \mathbb{Z}\right\} \) on the fundus image are recorded through an eye-tracker under the screen. Considering the natural field of view of human, we adapt a Gaussian filter function to enlarge the range of each fixation point, which is shown as
\begin{equation}
	G(x,y) =  \frac{1}{2\pi\sigma^{2}}e^{\frac{-x^{2}-y^{2}}{2\sigma^{2}}}
	\label{eq_1}
\end{equation}
where $\sigma$ is the variance of $G(x,y)$ mainly controlling the amplification range. According to the experimental experience of eye tracker, $\sigma$ is set to 60 pixels in our study. Finally, at the end of the reading stage, we obtain a gaze map that reflects the distribution of the ophthalmologist's attention.

In the reasoning stage, ophthalmologists usually conduct analysis on what they see on the fundus image. In general, the image regions with high ophthalmologists’ attention will become the main basis for diagnosis decisions. Referring to the reasoning process of the clinician, we select top $K$ local image patches with the size of $M \times M$ pixels in descending order of ophthalmologists’ attention on the gaze map. Notably, $K$ and $M$ are hyper-parameters that can be set according to different target diseases. To make interpretable and accurate reasoning, we imitate the characteristics of human reasoning and propose a novel artificial intelligence aided diagnosis model, dual-cross-attention MIL (DCAMIL) network, which will be described in the following sections.

\subsection{Problem Formulation}
We consider a multi-modal fundus image dataset $\mathbb{D}=\{(S_i, G_i, Y_i)\}_{i=1}^N$, where $S_i$ represents a color fundus image and $G_i$ is the corresponding gaze map of $S_i$. $Y_i \in \{0,1\}$ is the diagnostic label of $S_i$ for dichotomous judgment of health and disease. According to the attention distribution on the $G_i$, we crop $K$ local image patches from $S_i$ to generate the instances of a image bag $S_i^{'}$, i.e., $S_i^{'}=\{s_1^{(i)},s_2^{(i)}, ..., s_K^{(i)}\}$ that represents the $i^{th}$ image bag. We assume that all the instances $s_k^{(i)} (k={1,2,...,K})$ share an identical label, i.e., $Y_i$. However, as ophthalmologists still pay attention to some non-abnormal regions of the fundus image, not all instances are negative or positive in a bag. As a consequence, some instances own wrong labels, which can be regarded as noisy samples adverse for model learning. It is not only a weak supervision problem, but also a noise-tolerant learning problem. To address this challenge, we employ a weakly-supervised learning framework, i.e., MIL, to make bag-level fundus disease prediction. The binary MIL paradigm can be formulated as (\ref{eq:eq_2}).
\begin{equation}
	Y_{i}=\left\{\begin{array}{l}
		0, \quad \text { if } \sum_{k=1}^K y_{i k}=0 \\
		1, \quad \text { otherswise }
	\end{array}\right.
    \label{eq:eq_2}
\end{equation}
where $y_{ik}=\{0,1\}$, for $k=\{1,..., K\}$, is the label of instance in the $i^{th}$ bag with label $Y_i$. Under this setting, the bag label $Y_i=1$ if there are at least one instance labeled positive (i.e., $\sum_{k=1}^K y_{i k} > 0$ ). In this work, our target is to learn a MIL-based robust model that can detect fundus diseases (i.e., DR and AMD) in the ROIs of a fundus image. 

\subsection{Bag generation and augmentation} 
Firstly, the collected gaze map $G_i$ by the eye-tracker is processed into a single channel image with pixel value range of $[0, 255]$. Due to the same image size of $S_i$ and $G_i$, the pixel value on the $G_i$ reflect the degree of concern of the corresponding pixel on the $S_i$. Secondly, we adapt a sliding window with the size of $M \times M$ to collect sub-regions on the $G_i$ according to the step size of $M/2$. Thus, we can capture the complete anomaly area as much as possible. Thirdly, we calculate the attention value of each sub-region and sort them from large to small. Then, we select the sub-regions with top $K$ attention value, and crop the corresponding sub-regions on the $S_i$ generating a bag $S_i^{'}$. Ideally, the order of instances in each bag should not affect the results of model. Thus, we employ the sequence augmentation (SA) module, i.e., dynamic random permutation, during the model training to keep the permutation invariance of a bag. More concretely, the instance permutation in a bag during each training epochs can be different. Accordingly, the model will learn more the characteristics of each instance rather than the whole structure of the bag.

\begin{figure*}[!t]
	\centering
	\includegraphics[width=2\columnwidth]{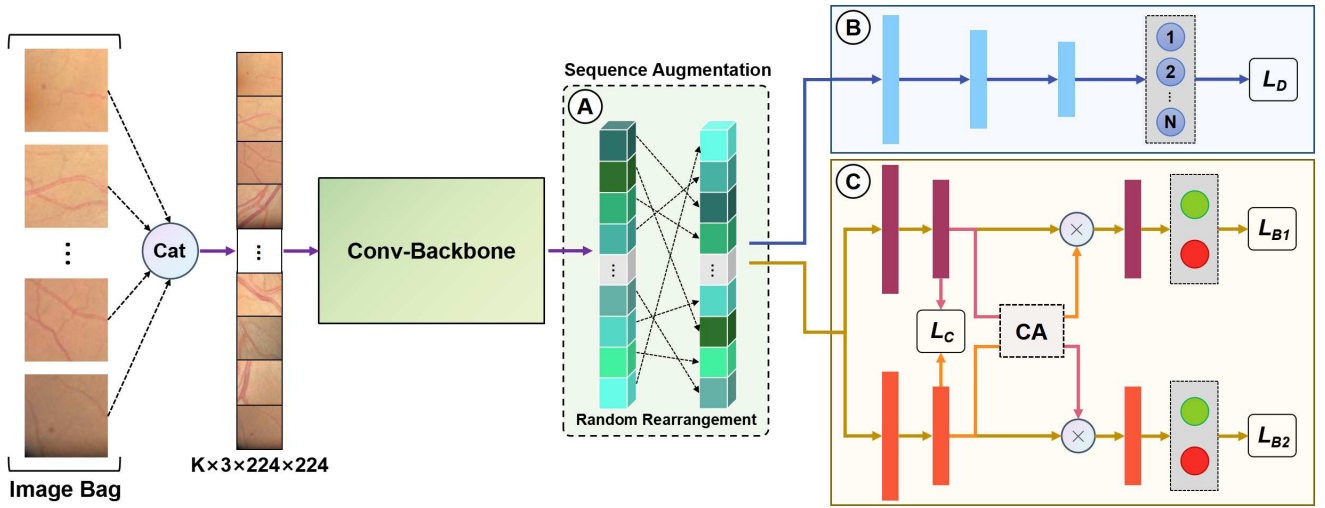}
	\vspace{-0.5em}
	\caption{The pipeline of our proposed method.}
	\label{fig:method}
	\vspace{1.0em}
\end{figure*}

\begin{figure}[!t]
	\centering
	\includegraphics[width=0.85\columnwidth]{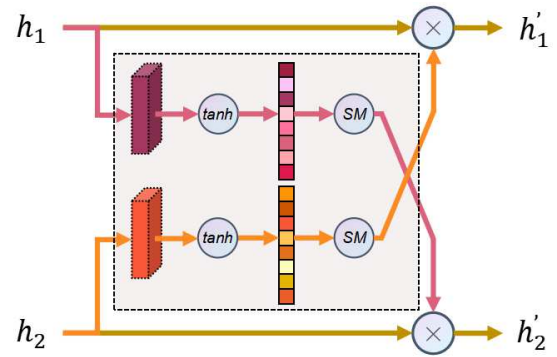}
	\vspace{-0.5em}
	\caption{Details of the CA module.}
	\label{fig:ca-module}
	\vspace{0em}
\end{figure}

\subsection{Daul-cross-attention network}
We rethink this weakly-supervised dilemma from the view of noise-tolerance learning, and innovatively propose a robust dual-cross-attention based architecture for MIL, i.e., DCAMIL network, which is illustrated in \textcolor{blue}{Fig \ref{fig:method}}. As confirmed by researchers, the learning process of DL model is similar to that of human. Specifically, the DL model gives priority to learning simple samples or obvious features, then learning difficult samples or inconspicuous features \cite{jiang2018mentornet}. With the training epochs increase, traditional MIL models will easily over-fit part of noisy instances losing robustness and generalization. The proposed DCAMIL network is composed of a backbone feature extractor, a dual-network-based multi-instance classifier and a domain adversarial network, which can both significantly relieve over-fitting phenomenon and enhance the anti-noise performance of fundus disease detection.

\subsubsection{Daul architecture} 
Since some dual-network structures have manifest strong anti-interference \cite{han2018co}, we take the dual-network as the multi-instance classifier after the the backbone feature extractor \( F_b(\cdot) \). Concretely, the \( F_b(\cdot) \) can be a typical CNN (e.g., ResNet or VGG) which maps a $224\times224$-pixel instance  $s_{i}^{\prime}(i \in[1, K])$ into a feature vector $g_i \in \mathrm{R}^Q$. Then, the corresponding feature vector of $S_{i}^{\prime}$ will be fed into a dual network that owns two individual bag classifiers, \( F_1(\cdot) \) and \( F_2(\cdot) \). Their predicted class probability vectors $P_{ij}(j \in\{1,2\})$ can be shown as \textcolor{blue}{Eq.\ref{eq_3}} and \textcolor{blue}{Eq.\ref{eq_4}}
\begin{equation}
	P_{i 1}=F_1\left(F_b\left(S_i^{\prime} \mid \theta_b\right) \mid \theta_1\right)
	\label{eq_3}
\end{equation}
\begin{equation}
	P_{i 2}=F_2\left(F_b\left(S_i^{\prime} \mid \theta_b\right) \mid \theta_2\right)
	\label{eq_4}
\end{equation}
where $\theta_1$, $\theta_2$ and $\theta_b$ represent the trainable parameters of $F_b$, $F_1$ and $F_2$, respectively. The optimization process of $F_{i}(i\in\{1,2\})$ is to minimize the cross-entropy loss function $L_{ce}$ between the $P_{ij}(j\in\{1,2\})$ and the true class label $Y_{i}$. We add the loss functions of two heads to form the classification loss function, which can be shown as \textcolor{blue}{Eq.\ref{eq_5}}
\begin{equation}
	L_1=\frac{1}{N} \sum_{i=1}^N\left(L_{c e}\left(Y_i, P_{i 1}\right)+L_{c e}\left(Y_i, P_{i 2}\right)\right)
	\label{eq_5}
\end{equation}
In particular, the parameter optimization of \( F_b(\cdot) \) is formulated as \textcolor{blue}{Eq.\ref{eq_6}}
\begin{equation}
	\theta_b^{\prime}=\theta_b-\frac{1}{N} \sum_{i=1}^N\left(\frac{\partial L\left(Y_i, P_{i 1}\right)}{\partial \theta_b}+\frac{\partial L\left(Y_i, P_{i 2}\right)}{\partial \theta_b}\right)
	\label{eq_6}
\end{equation}
The optimization direction of $\theta_b$ is towards the optimal points of both $L\left(Y_i, P_{i 1}\right)$ and $L\left(Y_i, P_{i 2}\right)$, which enhance the robustness and generalization of \( F_b(\cdot) \). 

\subsubsection{Cross attention module}
As the research of \cite{ilse2018attention} showed that the attention module can help the DL network focus on the important instances with diagnostic information. Thus, as each branch of the dual-network is adequately trained, they will have their own instances of interest. In our study, the attention module is regarded as an “instance filter” that can effectively suppress the influence of noisy instances. In order to make these two branch networks complement each other, we propose a cross-attention (CA) module that can establish an attention information interaction path. The basic purpose of the CA module is to achieve sample selection just like the Co-teaching method \cite{han2018co}, but differently the CA highlight the key instance through attention rather than preset noise rates, which is illustrated in \textcolor{blue}{Fig X}. The detailed architecture of CA module is shown in \textcolor{blue}{Fig \ref{fig:ca-module}}, and the attention interaction of dual-network is realized through following formula:
\begin{equation}
	h_1^{\prime}=h_1 \cdot \operatorname{softmax}\left(\tanh \left(h_2 \cdot W_{21}\right) \cdot W_{22}\right)
	\label{eq_7}
\end{equation}
\begin{equation}
	h_2^{\prime}=h_2 \cdot \operatorname{softmax}\left(\tanh \left(h_1 \cdot W_{11}\right) \cdot W_{12}\right)
	\label{eq_8}
\end{equation}

\subsubsection{Contrastive learning} 
Generally, noisy instances in the bag not only prevent the model from converging to a local optimum, but also destroy the intrinsic similarity of the model. Contrastive learning has been used to keep the intrinsic similarity in a single network through strong data transformation \cite{tan2021co}. In dual network structure, different initialization parameters lead two different branch networks having different feature mapping spaces for each sample, which also can be considered as two types of data transformations. To keep the intrinsic similarity of our dual-network, we employed the contrastive learning between the feature vectors before the CA module in each branch network. The intrinsic similarity loss in the dual-network is expressed as $L_2$ in formula \textcolor{blue}{Eq.\ref{eq_10}}.

\begin{footnotesize}
\begin{equation}
	L\left(h_1^{(i)}, h_2^{(i)}\right)=-\log \frac{\exp \left(d\left(h_1^{(i)}, h_2^{(i)}\right) / \tau\right)}{\sum_{j=1, i \neq j}^K \sum_{t_i, t_j \in\{1,2\}} \exp \left(d\left(h_1^{(i)}, h_2^{(i)}\right) / \tau\right)}
	\label{eq_9}
\end{equation}
\end{footnotesize}
\begin{equation}
	L_2=\sum_{i=1}^K L\left(h_1^{(i)}, h_2^{(i)}\right)
	\label{eq_10}
\end{equation}

\subsection{Domain adversarial network}
The style of color fundus photograph can vary greatly due to individual differences in patients and the impact of photographic equipment. To alleviate this problem, we introduced a domain predictor \( F_d(\cdot) \) after \( F_b(\cdot) \) shared with both \( F_1(\cdot) \) and \( F_2(\cdot) \), and implemented domain-adversarial (DA) \cite{ganin2016domain} training to ignore the style difference between bags. The DA module consists of three-layer neural network and the domain adversarial loss is formulated as:
\begin{equation}
	L_3=-\sum_{n=1}^N \frac{1}{K} \sum_{i=1}^K L_{c e}\left(D_n, F_d\left(g_i ; \theta_{D A}\right)\right)
\end{equation}
where $g_i$ is the $i^{th}$ feature vector of instance in the bag, $D_n \in\{1,2, \ldots, N\}$ represents the domain label of each bag and $\theta_{DA}$ expresses the trainable parameters of DA module.

As a whole, the proposed model is optimized by an integrated loss function as follows:
\begin{equation}
	L_{\text {total }}=\alpha L_1+\beta L_2+\gamma L_3
\end{equation}

\subsection{Evaluation metrics and comparative study}
\textbf{Evaluation metrics.} The performance is quantitatively evaluated by the commonly-used metrics, including the best test accuracy, precision, recall and F1 Score. As both the DR detection and AMD detection are binary classification tasks, we furthermore plot the receiver operator characteristic (ROC) curve and calculate the area under the ROC curve (AUC) for each approach. The evaluation metrics are defined as follows:

\begin{equation}
	Accuracy = \frac{TP+TN}{TP+FP+TN+FN}
	\label{eq_21}
\end{equation}
\begin{equation}
	Precision = \dfrac{TP}{TP+FP}
	\label{eq_22}
\end{equation}
\begin{equation}
	Recall = \dfrac{TP}{TP+FN}
	\label{eq_23}
\end{equation}
\begin{equation}
	F1 \; Score = 2*\dfrac{Precision*Recall}{Precision + Recall}
	\label{eq_24}
\end{equation}

Inside, TP, FP, TN and FN represent ``True Positive'', ``False Positive'', ``True Negative'' and ``False Negative'', respectively.

\textbf{Comparative study methods.} We compared the proposed DCAMIL method with other six popular MIL algorithms, including AB-MIL, CLAM, DSMIL, LA-MIL, MIL-VT, and MS-DA-MIL. To eliminate the specificity of different network architectures, we reimplemented with the same network backbone architecture (i.e., ResNet18) on all methods. Although these comparison methods were initially designed and evaluated for the whole slide image classification tasks, we readapted them with finetune hyperparameters on unified target fundus image datasets for a fair comparison.
%The compared methods include:
%\begin{enumerate}
%	\item \textbf{AB-MIL:} This is the common training procedure without any processing strategy {of} the noisy labels.  
%	\item \textbf{CLAM:} This is the method proposed by. 
%	\item \textbf{DSMIL:} This is the method proposed by .
%	\item \textbf{LA-MIL:} This method was to train .  
%	\item \textbf{MIL-VT:} This anti-noise method proposed a . 
%	\item \textbf{MS-DA-MIL:} This anti-noise method proposed a . 
%\end{enumerate}

\section{Experiment}
\subsection{Dataset construction}
\textbf{AMD-Gaze} dataset is composed of gaze maps of healthy and AMD fundus images during AMD diagnosis, and the diagnostic fundus images are from the private Fundus-iSee dataset (i.e., iSee-NORM and iSee-AMD). The gaze map collection process is just the AMD diagnostic process of ophthalmologists. We finally obtain 1097 gaze maps, including 557 healthy and 540 AMD gaze maps. 

\textbf{DR-Gaze} dataset contains gaze maps of DR diagnosis based on the publicly available DR dataset that is provided by EyePACS \cite{EyePACS}. In the source DR dataset, there are five rated levels ($0\sim4$), i.e., healthy (DR0), mild (DR1), moderate (DR2), severe (DR3) and proliferative DR (DR4). Among them, early DR detection, that is discrimination between DR0 and DR1, is quite significant and challenging. Thus, we simulate a early DR detection to accumulate ophthalmologists’ gaze maps. A total of 1020 gaze maps, including 498 DR0 and 522 DR1 gaze maps, are collected in the DR-Gaze dataset. 

The quantity distribution of training, validation and testing data on the AMD-Gaze and DR-Gaze datasets is shown in \textcolor{blue}{Table \ref{tab: data}}.
%Besides, the illustrative examples of DR-Gaze and AMD-Gaze can be shown in \textcolor{blue}{Fig X}.

\begin{table}[t!]
	\caption{The ablation study of each strategy. The best results are highlighted.}
	\begin{center}
		\label{tab: data}
		\vspace{-0.8em}
		\resizebox{0.85 \columnwidth}{13mm}{
			\begin{tabular}{ccccc} 
				\toprule[1.5pt]				  
				Dataset & Category & Train &	Validation	& Test  \\
				\midrule[0.75pt]
				\multirow{2}{*}{AMD-Gaze} & Negative & 357 & 100 & 100  \\
				& Positive & 340 & 100 & 100  \\
				\midrule[0.75pt]
				\multirow{2}{*}{DR-Gaze} & Negative & 299 & 100 & 100  \\
				& Positive & 321 & 100 & 100  \\				 	
				\bottomrule[1.5pt]
			\end{tabular}
		}
	\end{center}
	\vspace{-2.0em}
\end{table}

\subsection{Experiment setting}
In our experiments, the pre-trained weights were employed to initialize the back-bone network for all the methods. For all experimental methods, a total of 100 epoches is conducted in the training phase. For our proposed method, we used stochastic gradient descent optimizer with the learning rate of 0.0001 and the momentum of 0.9. Notably, the cropped image size of the instance from the AMD and DR images is $100\times100$ and $200\times200$, respectively, and then both the AMD and DR instances were uniformly resized into $224\times224$. The proposed model was implemented in Pytorch and all experiments were performed on two NVIDIA TITAN RTX GPUs of 24G memory.

\begin{figure}[!t]
	\centering
	\includegraphics[width=1\columnwidth]{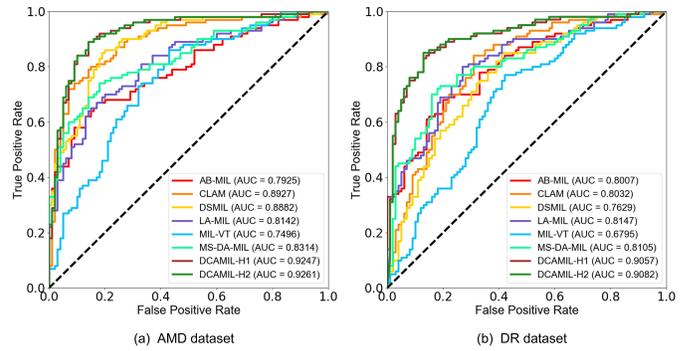}
	\vspace{-0.5em}
	\caption{ROC curves of the representative classification methods with multiple instance learning on (a) AMD dataset  and (b) DR dataset.}
	\label{fig:ROC}
	\vspace{0.0em}
\end{figure}

\begin{table}[t!]
	\caption{Comparison of different methods based on AMD dataset.}
	\begin{center}
		\label{tab: comparison_AMD}
		\vspace{-2.2em}
		\resizebox{1 \columnwidth}{16.0mm}{
			\begin{tabular}{cccccc} 
				\toprule[1.5pt] 
				Methods & Accuracy & Recall & Precision & F1 Score & AUC  \\ 
				\midrule[0.75pt] 	
				AB-MIL & 0.73 & 0.54 & 0.8571 & 0.6626 & 0.7925\\ 
				CLAM & 0.81 & 0.76 & 0.8440 & 0.8000 & 0.8927\\ 
				DSMIL & 0.82 & 0.77 & 0.8556 & 0.8152 & 0.8940\\ 
				LA-MIL & 0.74 & 0.65 & 0.8025 & 0.7182 & 0.8142\\ 
				MIL-VT & 0.70 & 0.73 & 0.6887 & 0.7087 & 0.7496\\ 
				MS-DA-MIL & 0.70 & 0.44 & \textbf{0.9167} & 0.5946 & 0.8314\\ 
				DCAMIL-H1 (Ours) & \textbf{0.86} & \textbf{0.92} & 0.8142 & \textbf{0.8639} & 0.9247\\ 
				DCAMIL-H2 (Ours) & 0.85 & 0.92 & 0.8070 & 0.8598 & \textbf{0.9261}\\ 
				\bottomrule[1.5pt]
			\end{tabular}
		}
	\end{center}
	\vspace{-0.5em}
\end{table}

\section{Results analysis}
\subsection{Evaluation on the AMD dataset}
According to the key instances extracted by the eye-tracking, we compared six representative MIL methods with the proposed DCAMIL on the AMD dataset. AMD disease is related to the location of lesions on the retina, that is, lesions may appear in or around the macular area of the fundus. Therefore, lesions in other locations on the retina will also attract ophthalmologist’s attention and become the main noise instances in the negative image bags. As the proposed DCAMIL model has two independent output heads, we reported both the results of the two heads. \textcolor{blue}{Fig \ref{fig:ROC}.(a)} shows ROC curves for all methods on the AMD test set, indicating that the proposed DCAMIL model is more robust than other comparative models. Moveover, DCAMIL-H1 and DCAMIL-H2 have approximate performance. \textcolor{blue}{Table \ref{tab: comparison_AMD}} further presents the comparison results of detailed evaluation metrics, which illustrates that our method surpass reference methods in most all quantitative indicators. 

\subsection{Evaluation on the DR dataset}
As for the early DR detection task, the related lesions may be found anywhere on the retina. In addition to noise negative instances in the positive image bag, most lesions are small and not very significant in the positive image bag, especially the MA. Hence, the difference between positive and negative instances is quite small, which increases the difficulty of early DR detection. Compared with state-of-the-arts, our model is better than other models by a large margin on the ROC curve, which is shown in \textcolor{blue}{Fig \ref{fig:ROC}.(b)}. As the DCAMIL has strong noise resistance and important feature capture ability, the performance indicators of early DR detection on the DR test set demonstrate that it is superior to other state-of-the-arts, especially in terms of accuracy, recall and AUC value. The detailed comparison of performance indicators is shown in \textcolor{blue}{Table \ref{tab: comparison_DR}}.

\begin{table}[t!]
	\caption{Comparison of different methods based on DR dataset.}
	\begin{center}
		\label{tab: comparison_DR}
		\vspace{-2.0em}
		\resizebox{1 \columnwidth}{16.0mm}{
			\begin{tabular}{ccccccc} 
				\toprule[1.5pt]  
				Methods & Accuracy & Recall & Precision & F1 Score & AUC  \\ 
				\midrule[0.75pt]				
				AB-MIL & 0.70 & 0.74 & 0.6916 & 0.7150 & 0.8007 \\ 
				CLAM & 0.76 & 0.84 & 0.7179 & 0.7741  & 0.8032 \\ 
				DSMIL & 0.60 & \textbf{0.99} & 0.5531 & 0.7097  & 0.7629 \\ 
				LA-MIL & 0.74 & 0.68 & 0.7816 & 0.7273  & 0.8147 \\ 
				MIL-VT & 0.65 & 0.67 & 0.6442 & 0.6568  & 0.6795 \\ 
				MS-DA-MIL & 0.75 & 0.73 & \textbf{0.7604} & 0.7449  & 0.8105 \\ 
				DCAMIL-H1 (Ours) & 0.80 & 0.92 & 0.7419 & 0.8214 & 0.9057 \\ 
				DCAMIL-H2 (Ours) & \textbf{0.80} & 0.92 & 0.7419 & \textbf{0.8214} & \textbf{0.9082} \\ 
				\bottomrule[1.5pt]
			\end{tabular}
		}
	\end{center}
	\vspace{0.0em}
\end{table}

\subsection{Stability analysis of model training}
\begin{figure}[!t]
	\centering
	\includegraphics[width=1\columnwidth]{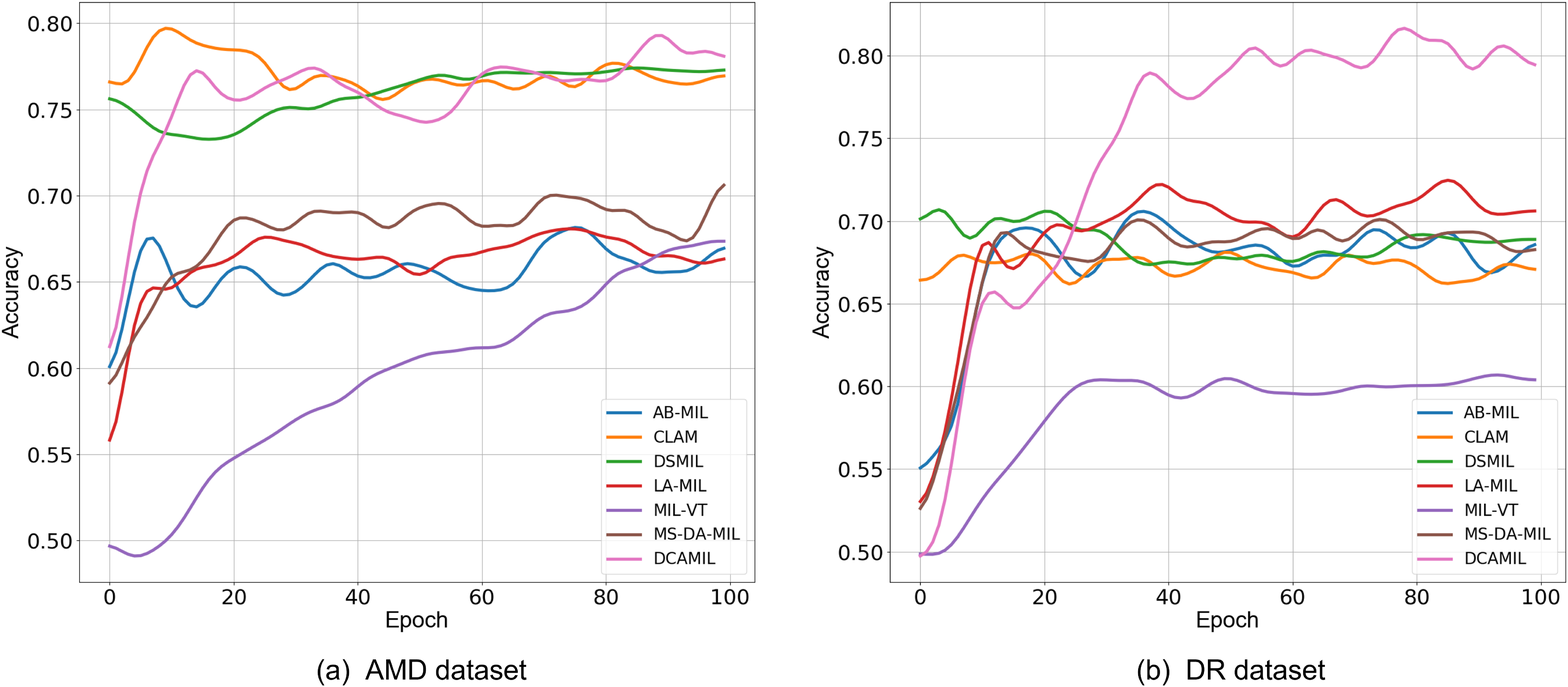}
	\vspace{-1.5em}
	\caption{Stability comparison of the representative MIL methods with validation accuracy against the training epoch. (a) and (b) are for the AMD dataset and DR dataset, respectively.}
	\label{fig:stable}
	\vspace{0.0em}
\end{figure}

We evaluated the accuracy changes of all the compared models on the validation sets of AMD and DR datasets during the training process, which is displayed in \textcolor{blue}{Fig \ref{fig:stable}}. As \textcolor{blue}{Fig \ref{fig:stable}.(a)} shows, the accuracy of the CLAM, DSMIL and DCAMIL methods on the AMD validation set is approximate, and our method manifests steady improvement on validation accuracy as the training epoch increases finally achieving the top performance. From \textcolor{blue}{Fig \ref{fig:stable}.(b)}, even though the lesions in early DR detection are more challenging than those in AMD detection, our method achieves an absolute leadership in validation accuracy on the DR dataset. By comparing the performance of our method on the validation and the test set of AMD and DR datasets, the proposed DCAMIL model has better stability and generalization ability.

\begin{table*}[b!]
	\caption{Test accuracy of different backbones. The best results are highlighted.}
	\begin{center}
		\label{tab: backbones}
		\vspace{-1.0em}
		\resizebox{1.95 \columnwidth}{14.5mm}{
			\begin{tabular}{cccccccccccc} 
				\toprule[1.5pt]
				\multirow{2}{*}{Backbones} & \multicolumn{5}{c}{AMD Dataset} & \multicolumn{6}{c}{DR Dataset} \\
				%\cline{2-6} \cline{8-12}
				\cmidrule[0.75pt]{2-6} \cmidrule[0.75pt]{8-12}
				&  Accuracy & Recall & Precision & F1 Score & AUC & & Accuracy & Recall & Precision & F1 Score & AUC \\
				\midrule[0.75pt]
				ResNet18 & \textbf{0.86} & 0.92 & 0.8142 & 0.8639 & \textbf{0.9247} & & 0.80 & \textbf{0.92} & 0.7419 & 0.8214 & 0.9082 \\
				ResNet50 & 0.82 & 0.84 & 0.8077 & 0.8235 & 0.8414 & & 0.78 & 0.83 & 0.7477 & 0.7867 & 0.8592 \\
				VGG16 & \textbf{0.86} & 0.87 & \textbf{0.8614} & \textbf{0.8657} & 0.9229 & & \textbf{0.88} & 0.88 & \textbf{0.8713} & \textbf{0.8756} & \textbf{0.9415} \\
				EfficientNet & 0.83 & \textbf{0.94} & 0.7705 & 0.8469 & 0.9209 & & 0.76 & 0.71 & 0.7978 & 0.7513 & 0.8355 \\            					 	
				\bottomrule[1.5pt]
			\end{tabular}
		}
	\end{center}
	\vspace{0.0em}
\end{table*}

\subsection{Evaluation on multiple embedding backbones}
We also validated the DCAMIL method on different popular CNN architectures that can be regarded as embedding backbones. \textcolor{blue}{Table \ref{tab: backbones}} demonstrates evaluation results of DCAMIL with four different backbones, including ResNet18, ResNet50, VGG16 and EfficientNet on two fundus disease datasets. Note that, the best performance among the two classifier heads from DCAMIL is reported for the final evaluation. From \textcolor{blue}{Table \ref{tab: backbones}}, we observe that the performance differences of DCAMIL method using different basic networks as backbone on the AMD dataset are not particularly significant. Specifically, almost the same optimal performance can be achieved using ResNet18 or VGG16 as the backbone network. However, different backbone networks present quite different performance of the early DR detection. In particular, taking VGG16 as the backbone network acquires the best performance on most evaluation metrics. For a fairness, we merely utilized ResNet18 as the same backbone in all the comparison experiments.

\begin{figure}[!t]
	\centering
	\includegraphics[width=1\columnwidth]{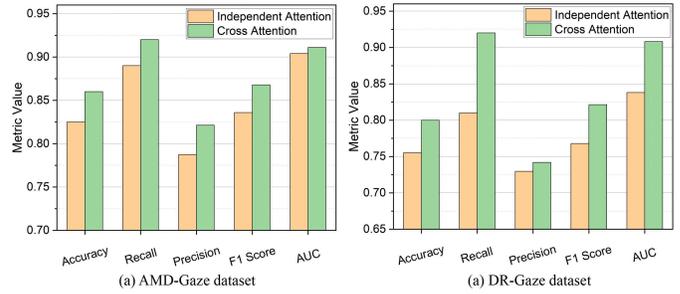}
	\vspace{-1.5em}
	\caption{Performance comparison of the proposed method with different attention maps. (a) and (b) are for AMD and DR dataset, respectively.}
	\label{fig:ablation_attention}
	\vspace{0.0em}
\end{figure}

\subsection{Comparison of attention methods}
We explore the effects of different attention maps, including independent attention and cross attention mechanisms. The comparison results in \textcolor{blue}{Fig \ref{fig:ablation_attention}} show that compared to independent attention, cross attention can comprehensively improve all performance metrics on the AMD-Gaze and DR-Gaze datasets, including accuracy, recall, percision, F1 Score and AUC. compared to independent attention, cross attention can comprehensively improve all performance indicators on the AMD-Gaze and DR-Gaze datasets. Theoretically, independent attention mechanism itself owns the ability to focus on important instances, but it dose not benefit from the dual-network thus to refer to the results of the peer branch network. Instead, the proposed cross attention can take advantage of the dual-network to further suppress noise instances and acquire better fundus disease detection results.

\begin{table*}[t!]
	\caption{Test accuracy of different modules. The best results are highlighted.}
	\begin{center}
		\label{tab: ablation_module}
		\vspace{-1.0em}
		\resizebox{1.95 \columnwidth}{14.0mm}{
			\begin{tabular}{cccccccccccc} 
				\toprule[1.5pt]
				\multirow{2}{*}{Methods} & \multicolumn{5}{c}{AMD-Gaze Dataset} & \multicolumn{6}{c}{DR-Gaze Dataset} \\
				\cmidrule[0.75pt]{2-6} \cmidrule[0.75pt]{8-12}
				&  Accuracy & Recall & Precision & F1 Score & AUC & & Accuracy & Recall & Precision & F1 Score & AUC \\
				\midrule[0.75pt]
				DCAMIL & \textbf{0.86} & 0.92 & 0.8142 & \textbf{0.8639} & 0.9247 & & \textbf{0.80} & 0.92 & \textbf{0.7419} & \textbf{0.8214} & \textbf{0.9082} \\
				DCAMIL-SA & 0.84 & 0.79 & \textbf{0.8876} & 0.8360 & 0.9058 & & 0.77 & \textbf{0.85} & 0.7328 & 0.7871 & 0.8409 \\
				DCAMIL-DA & 0.82 & \textbf{0.95} & 0.7540 & 0.8407 & \textbf{0.9303} & & 0.76 & 0.84 & 0.7241 & 0.7778 & 0.8782 \\
				DCAMIL-SA-DA & 0.82 & 0.79 & 0.8495 & 0.8187 & 0.9057 & & 0.72 & 0.75 & 0.7075 & 0.7281 & 0.8170\\ 
				\bottomrule[1.5pt]
			\end{tabular}
		}
	\end{center}
	\vspace{0.0em}
\end{table*}

\begin{figure*}[!b]
	\centering
	\includegraphics[width=2.0\columnwidth]{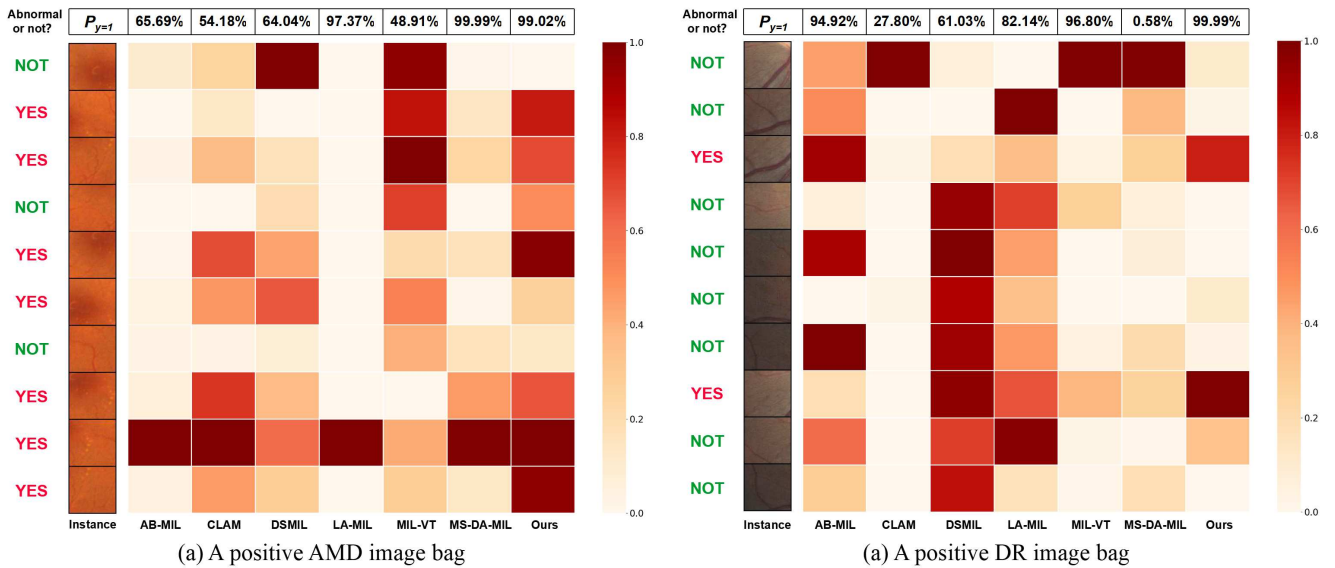}
	\vspace{0.0em}
	\caption{Illustrative examples of visualization analysis for the positive AMD and DR case which indicates that ours can cherry-pick diagnostic related instances, facilitating the classification performance.}
	\label{fig:instance-result}
	\vspace{1.7em}
\end{figure*}

\subsection{Ablation study}
\textbf{Ablation study on each key technology.} We conducted ablation study on each key technology of the DCAMIL model, including dual-network (DN), contrastive learning (CL) and cross attention (CA). In these experiments, we kept sequence augmentation (SA) and domain-adversial (DA) modules in the model. Notably, the CL only works on the DN architecture. From \textcolor{blue}{Table \ref{tab: ablation_strategy}}, each new-adding technology improve the performance of fundus disease detection on both accuracy and F1 Score to some extent. For the early DR detection, integrating the three technologies together achieves the best performance on all indicators. Among them, the CA played the greatest role.

\textbf{Ablation study on each data transform module.} Furthermore, we conduct ablation study on two different data transform modules, SA and DA, which is shown in \textcolor{blue}{Table \ref{tab: ablation_module}}. The role of SA module is to enrich the diversity of image bags and the main function of DA module is to eliminate the influence of unique style of different color fundus photography. In these ablation experiments, we retained the DN, CL and CA technologies in our model. The experimental results prove that both the SA and DA module enchance the performance of AMD and early DR detection.

\begin{table}[t!]
	\caption{The ablation study of each strategy. The best results are highlighted.}
	\begin{center}
		\label{tab: ablation_strategy}
		\vspace{-1.0em}
		\resizebox{0.99 \columnwidth}{15mm}{
			\begin{tabular}{ccccccccc} 
				\toprule[1.5pt]				  
				Dataset &	DN &	CL	& CA &	Accuracy	& Recall & Precision &	F1 Score & AUC \\
				\midrule[0.75pt]
				\multirow{4}{*}{AMD-Gaze} & $\times$ & $\times$ & $\times$ & 0.84 & 0.80 & \textbf{0.8696} & 0.8333 & 0.9186  \\
				& $\checkmark$ & $\times$ & $\times$ & 0.84 & \textbf{0.95} & 0.7724 & 0.8520 & \textbf{0.9268} \\
				& $\checkmark$ & $\checkmark$ & $\times$ & 0.85 & 0.88 & 0.8302 & 0.8544 & 0.9253  \\
				& $\checkmark$ & $\checkmark$ & $\checkmark$ & \textbf{0.86} & 0.92 & 0.8142 & \textbf{0.8639} & 0.9247 \\
				\midrule[0.75pt]
				\multirow{4}{*}{DR-Gaze} & $\times$ & $\times$ & $\times$ & 0.72 & 0.74 & 0.7048 & 0.7220 & 0.7815  \\
				& $\checkmark$ & $\times$ & $\times$ & 0.74 & 0.76 & 0.7308 & 0.7451 & 0.8163  \\
				& $\checkmark$ & $\checkmark$ & $\times$ & 0.75 & 0.80 & 0.7273 & 0.7619 & 0.8355  \\
				& $\checkmark$ & $\checkmark$ & $\checkmark$ & \textbf{0.80} & \textbf{0.92} & \textbf{0.7419} & \textbf{0.8214} & \textbf{0.9082} \\ 						 	
				\bottomrule[1.5pt]
			\end{tabular}
		}
	\end{center}
	\vspace{0.0em}
\end{table}

\subsection{Visualization Analysis}
To demonstrate the interpretability of our method, we presented the prediction effects of image bags on the AMD-Gaze and DR-Gaze datasets through illustrative examples, which is illustrated in \textcolor{blue}{Fig \ref{fig:instance-result}}. \textcolor{blue}{Fig \ref{fig:instance-result}.(a)} and \textcolor{blue}{Fig \ref{fig:instance-result}.(b)} shows the visualization results of a positive AMD image bag and a positive DR image bag, respectively. The vertical column of the color matrix represents the attention degree of each method to each instance in the image bag. For the AMD image bag, more than half of the instances are positive, our method gives more attention to thoes positive instances thus to present more reasonable intergretability. For the positive DR image bag with many negative instances, our method accurately captures the positive instances, demonstrating the excellent noise resistance of our model. Meanwhile, our method outputs quite high predictive confidence for both the AMD and DR image bags.

\section{Discussion}
\subsection{Number of instances in the image bag}
\begin{figure}[!t]
	\centering
	\includegraphics[width=1\columnwidth]{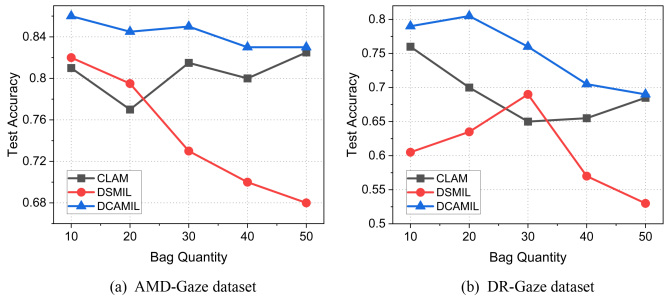}
	\vspace{-1.5em}
	\caption{Study of the impact of bag quantity on the classification performance. (a) and (b) are for AMD dataset and DR dataset, respectively.}
	\label{fig:bag}
	\vspace{1.0em}
\end{figure}

As mentioned in the previous section, we selected the top $K$ $(K=10)$ ROIs as instances of a image bag with the help of eye-tracking. There is a default fact in this study that the more attention an ophthalmologist assigns to an ROI, the more diagnostic information it contains. Hence, we investigated the influence of the value of K, i.e., the number of instances in the image bag, for the AMD and DR detection. We chosed the methods that achieved the second best results on the AMD-Gaze and DR-Gaze datasets in \textcolor{blue}{Table \ref{tab: comparison_AMD}} and  \textcolor{blue}{Table \ref{tab: comparison_DR}} (i.e., DSMIL and CLAM), respectively, for comparative analysis. Concretely, we analyzed the performance of three methods at $K=\{10, 20, 30, 40, 50\}$ that is shown in \textcolor{blue}{Fig \ref{fig:bag}}. For both the AMD and early DR detection, the proposed DCAMIL achieved optimal performance on all $K$ values. As the $K$ value increases, the overall performance trend of both the DCAMIL and DSMIL models decreases, although there is a rebound in some $K$ values. Notably, the test accuracy of the CLAM model exhibits significant fluctuations and reaches its highest at $K=50$ on the AMD-Gaze dataset. Although a large $K$ is beneficial for the cluster-based CLAM method, its weak noise resistance still offsets the performance improvement.

\subsection{Comparison of two instance generation ways}
\begin{figure}[!t]
	\centering
	\includegraphics[width=1\columnwidth]{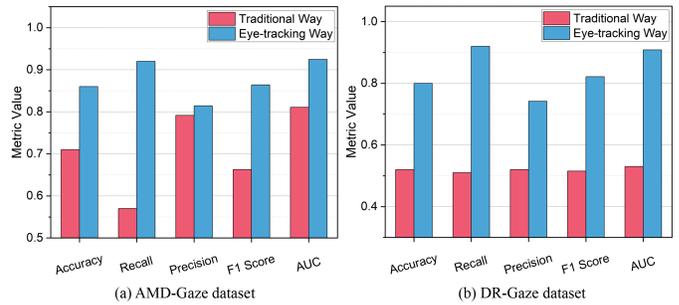}
	\vspace{-1.5em}
	\caption{Performance comparison of the proposed method with different attention maps. (a) and (b) are for AMD and DR dataset, respectively.}
	\label{fig:instance-gen-method}
	\vspace{0.0em}
\end{figure}

One of the novel points in this study is that we employed the eye-tracking information of ophthalmologists during their diagnosis to extract ROIs and then proposed the DCAMIL model to finally construct a HITL CAD for fundus disease detection. However, we would like to know how significant eye-tracking information is for fundus disease detection. Thus, we compared the performance metrics between the eye-tracking way and traditional way to generate the instances for the image bag, which can be seen in \textcolor{blue}{Fig \ref{fig:instance-gen-method}}. The comparison results show that the eye-tracking way outperforms the traditional way in all indicators. In particular, the traditional way for the DCAMIL model is almost completely ineffective in the early DR detection, further demonstrating the importance of the eye-tracking way for the DCAMIL model.

%\subsection{Model complexity and time cost analysis}
%We further conduct   

\iffalse
\begin{table}[t!]
	\caption{Model complexity and time cost comparison.}
	\begin{center}
		\label{tab: Model_size_time}
		\vspace{-1.0em}
		\resizebox{0.8 \columnwidth}{16mm}{
			\begin{tabular}{ccc} 
				\toprule[1.5pt]
				Methods & Model size(M) & Inference time(ms) \\ 
				\midrule[0.75pt]				
				AB-MIL & 242.98 & 9.6654    \\
				CLAM & 46.62 &  0.8110   \\
				DSMIL & 44.97 & 41.3913     \\
				LA-MIL & 242.98 & 9.0194     \\
				MIL-VT & 125.76 &  12.0519    \\
				MS-DA-MIL & 242.98 &  9.3692    \\
				DCAMIL (Ours) & 443.15 &  9.7437    \\ 
				\bottomrule[1.5pt]
			\end{tabular}
		}
	\end{center}
	\vspace{-1.0em}
\end{table}
\fi

\section{Conclusion}
In this study, we introduced an eye-tracker during the disease diagnosis process of color fundus photography by ophthalmologists. Concretely, AMD and early DR are respectively taken as target diagnostic diseases. We innovatively validated the feasibility of a HITL CAD system based on eye-tracking and proposed a robust DCAMIL model for AMD and early DR detection. The experimental results demonstrate that the eye-tracking information from ophthalmologists can help enhance diagnostic performance. Additionally, the proposed DCAMIL model can achieve the absolutely competitive performance on the HITL CAD for the AMD and early DR diseases. Courrently, data-driven CAD algorithms for fundus diseases aret the majority, but we must not overlook the value of medical prior knowledge, such as medical diagnostic knowledge contained in ophthalmologists' eye-tracking or gaze maps.

\bibliographystyle{IEEEtran}%IEEEtran ieeecolor
\bibliography{refs}

\end{document}